# Microelectromechanical components in electrical metrology

**Antti MANNINEN[1], Anu KÄRKKÄINEN[2], Nadine PESONEN[2], Aarne OJA[2], and Heikki SEPPÄ[2]**

[1]Centre for Metrology and Accreditation (MIKES), P.O. Box 9, FI-02151 Espoo, Finland
[2]VTT, P.O. Box 1207, FI-02044 VTT, Finland

**Abstract**

*Microelectromechanical systems (MEMS) can offer a competitive alternative for conventional technology in electrical precision measurements. This article summarises recent work in development of MEMS solutions for electrical metrology. MEMS-based voltage references, RMS-to-DC converters, high frequency power sensors, and reference oscillators are discussed. The main principle of operation of the components is the balance between electrical forces and mechanical spring forces in micromachined silicon structures. In RMS sensors and RMS-to-DC converters, the quadratic voltage dependence of the force between plates of a moving-plate capacitor is utilised, and the operation of the MEMS voltage reference is based on the pull-in phenomenon of a moving-plate capacitor. Advantages of MEMS devices compared to more conventional solutions include small size, low power consumption, low price in mass production, stability, and low 1/f noise. The drift caused by electrostatic charging effects has turned out to be a major problem. This problem has not yet been solved in DC applications, but it can be circumvented by using AC actuation instead of DC and by compensating the internal DC voltages of the component. In this way, an AC voltage reference with relative drift rate below $2\cdot10^{-6}$ during a three-week test period has been constructed. Even better stability has been demonstrated with a MEMS-based reference oscillator: no changes in resonance frequency were observed at relative uncertainty level of about $10^{-8}$ in a measurement which was continued for more than a month. MEMS components have also been developed for measuring RF and microwave power up to frequencies of about 40 GHz. Unlike conventional high frequency power sensors, which measure the absorbed power, the MEMS device measures the power that is transmitted through the sensor.*

## 1  Introduction

Importance of microelectromechanical systems (MEMS), or microsystem technology (MST), is continuously increasing. For example, a modern car contains about 20 - 40 MEMS-based accelerometers and other sensors. In addition to automotive and transportation industry, main applications of MEMS components are currently found in communication technology, medical and biological technologies, environmental control, etc. Hard disc drive read/write heads, inkjet printer heads, accelerometers and pressure sensors are well known mass market applications. In this article, we discuss applications of micromachined MEMS components in electrical precision measurements or, more specifically, in electrical metrology.

MEMS components are characterised by their small size (characteristic dimensions typically between 1 µm and 1 mm), low cost, low power consumption, reliability, and integrability with electronics. Typical MEMS products are mechanical actuators, sensors and other devices, which are microscopic in scale and are fabricated by integrated circuit technology. The operation principle of these components can be based on various physical phenomena, such as mechanical or thermal properties of micromachined structures, and the components often include integrated microelectronic circuitry.

The metrological applications discussed in this article are based on components which have, in principle, a similar moving-plate capacitor structure as MEMS accelerometers or pressure sensors. The components exploit movable micromachined beams, membranes, or cantilevers. When either DC or AC voltage is applied across the moving-plate capacitor, the balance between electrical and mechanical forces gives rise to two properties which make the component suitable for electrical metrology. First, electrical force is proportional to the square of the applied voltage, which means that the component can be used as a true RMS (root mean square) sensor or as an AC/DC converter. Second, the voltage across the component has a characteristic maximum, so-called pull-in voltage, which depends only on material properties and geometry of the component, and can be used as a stable voltage reference.

The potential of MEMS in electrical metrology has been earlier summarised by Seppä et al. [1] and by Wolffenbuttel and van Mullem [2]. The first suggested application of a micromechanical moving-plate capacitor structure in electrical metrology was the RMS-to-DC converter, whose principle of operation was published in 1995 [3]. Later it was proposed that a similar principle can be used for measuring RF and microwave power, as well [1]. Application of the pull-in voltage of a moving-plate capacitor as a DC or AC voltage reference was first proposed in 1998 [4].

Moving-plate MEMS devices have a great potential in electrical metrology and instrumentation, but a lot of intensive research is still required before actual instruments or standards will be commercially available. However, operation of many different devices has been demonstrated already. These include RMS-to-DC converters fabricated using both surface micromachining [5] and bulk micromachining [6]. An RMS-to-DC converter based on a SOI (silicon-on-insulator) process is presented in Ref. [7]. The first experimental results of pull-in voltage as a voltage reference were published in 2001 [1], [8], [9]. Between autumn 2001 and spring 2005, several European institutes and companies (VTT, MIKES, and VTI Technologies Ltd from Finland, Twente University and NMi-VSL from the Netherlands, PTB from Germany, and Fluke PM from Great Britain) developed MEMS devices for electrical metrology in a EU-funded collaboration project EMMA, which was coordinated by VTT. Main results of that project were fabrication and design of MEMS-based microwave power sensors up to frequencies of about 40 GHz [10], [11], and development of an AC voltage reference which was demonstrated to be stable within the 2-ppm (parts per million) experimental accuracy during the 3-week measurement period [12], [13]. This article is mostly based on results obtained in the EMMA project.

# 2 Basics of MEMS for electrical metrology

## 2.1 MEMS components

### 2.1.1 Manufacturing technologies

Silicon micromechanics is traditionally divided into two categories: surface micromechanics and bulk micromechanics [14]. The silicon-on-insulator (SOI) process falls between surface and bulk micromachining as far as the thickness of the structures and the lateral resolution are concerned. The unconventional surface micromachining technique LIGA (Litographie Galvanoformung, Abformung) is not discussed here.

Bulk micromachining is mainly used in manufacturing large seismic mass devices which have applications for example in high precision, low-g accelerometers. Often the whole wafer thicknes (typically about 0.4 mm) is used to process the moving beam/cantilever and the cavity is formed by bonding this wafer with another silicon or glass wafer. Bulk micromachining is based on etching single-crystal silicon either chemically (wet etching) or using plasma etching or reactive ion etching (RIE) techniques.

Surface micromachining is a mature processing technology generally used in manufacturing of high-g accelerometers, pressure sensors, etc. In surface micromachining the devices are made of thin films deposited on the surface of a silicon wafer or another substrate, which acts merely as a support for the structure. The most widely used materials are low-pressure chemical vapour deposited (LPCVD) polycrystalline silicon as the structural material and silicon dioxide as the sacrificial layer. The LPCVD polycrystalline layers are typically a few micrometers thick. The major advantages of surface micromachining are the simplicity of the process and the use of standard CMOS fabrication. The major challenge is the internal stress control of the polysilicon film, which is originally under compressive stress, and the variation of the stress along the growth direction. Also, planarity and stiction problems are common in surface micromachining.

The SOI process is based on a silicon-on-insulator (SOI) wafer, which is composed of two single-crystal silicon layers with silicon oxide in between. The relatively thin (1 – 2 µm) oxide layer can be used as an etch stop, an isolation layer, or as a sacrificial layer. Hence, relatively thick silicon structures can be realised with well defined heights. A wide range of thicknesses, from about 100 nm to about 1 mm, are available for the device layer of the SOI wafer. The internal stress in SOI wafers is low. Prosessing phases of a SOI component developed for the voltage reference are presented in Fig. 1.

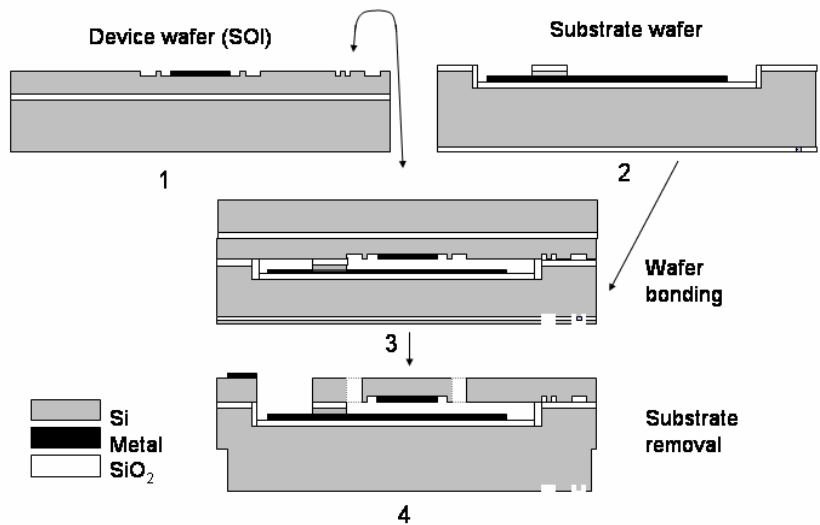

Figure 1 - Manufacturing steps of a SOI component [15]. 1) Processing of the device wafer (which contains the moving plate of the capacitor). 2) Processing of the substrate wafer (with the fixed electrode). 3) Bonding of the wafers. 4) Processing of the bonded pair.

### 2.1.2 Moving-plate capacitor structure

Figure 2a shows schematically the basic principle of a micromechanical moving-plate capacitor which is suitable for electrical metrology. It consists of two electrodes, one of which is suspended with a spring, for example a silicon beam which has the required compliancy determined by its geometry. Voltage $V$ applied across the capacitor creates an electrostatic force $F_{el}$ between the capacitor plates. The elastically suspended electrode is displaced until the restoring spring force $F = kx$ balances $F_{el}$.

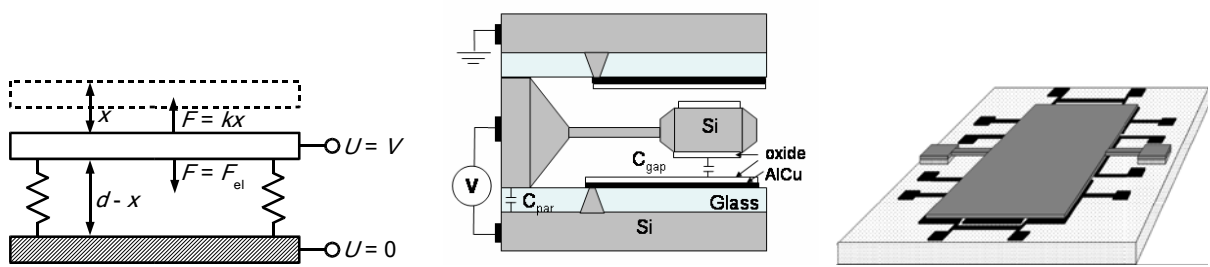

Figure 2 - a) Principle of moving-plate capacitor structure. b) MEMS accelerometer designed and fabricated by VTI Technologies Ltd [16]. c) Seesaw type SOI component developed for voltage reference applications by VTT [17].

Schematic views of two practical realisations of the moving-plate MEMS capacitor are presented in Figs. 2b and 2c. The component of Fig. 2b is a dice from a bulk micromachined stack of three silicon wafers and has outer dimensions of about $(1.5 \text{ mm})^3$. The boron-doped single crystal silicon wafer, from which the moving cantilever in the middle has been etched, is anodically bonded between two glass-covered silicon wafers. The component has been designed for acceleration measurements. Being sensitive to vibrations and inclination angle, it is not optimal for use in electrical metrology. However, it has given the best measurement results as a voltage reference so far. The component of Fig. 2c has been manufactured using SOI technology (see Fig. 1). To minimise the effect of mechanical vibrations, it has a seesaw structure with single crystal silicon as a torsional spring. The dimensions of the seesaw plate are 550 × 800 × 20 µm$^3$.

Single-crystal silicon, which is used as the structural material of the bulk-micromachined and SOI-based structures of Fig. 2, is ideal for metrological applications due to its excellent mechanical properties and stability. In surface-micromachined structures, polycrystalline silicon (polysilicon) or other materials are used, instead, because it is difficult to grow thin films of single-crystal silicon. The electrode surfaces are either strongly doped or metal-coated to decrease slow electrostatic charging effects, which are a major factor limiting the stability of the MEMS components. Common coating materials are aluminium (Al), molybdenum (Mo), tungsten (W), copper (Cu), and tantalum (Ta), and several compounds and alloys, such as TiW and Al-Cu.

## 2.2  Principle of operation

Consider the simplified moving-plate capacitor of Fig. 2a. Voltage $V$ applied across the capacitor plates causes an attractive force

$$F_{el,V} = \frac{\varepsilon A}{2(d-x)^2} V^2, \tag{1}$$

where we have assumed parallel-plate capacitance $C(x) = \varepsilon A/(d-x)$. Here $\varepsilon$ is the permittivity of the medium, $A$ is the electrode area, and $d$ is the gap between the electrodes when $V = 0$. This force is balanced by the restoring spring force $F = -kx$, and the relation between voltage $V$ and displacement $x$ becomes

$$V = (d-x)\sqrt{x \cdot 2k/(\varepsilon A)} \quad . \tag{2}$$

This relation is shown graphically in Fig. 3.

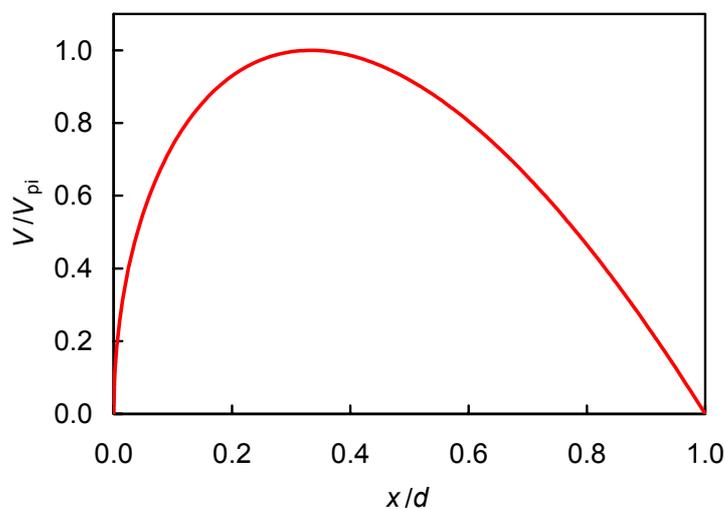

Figure 3 -   Voltage of a moving-plate capacitor as a function of displacement. See Eq. (3) for the definition of the pull-in voltage $V_{pi}$.

In deriving Eq. (2) we have assumed static displacement $x$, i.e. that the component does not oscillate mechanically due to the applied voltage. This is naturally valid for DC. In the case of AC voltage, displacement is static if the frequency of the applied voltage is much higher than the mechanical resonance frequency of the component or if the oscillations are damped for example by gas between the plates. In this case, the component works as a true RMS-to-DC converter: because $F_{el} \propto V^2$, the average force (and, thus, the displacement of the moving electrode) caused by AC voltage $V(t)$ and DC voltage $V_{DC}$ are equal if time-average of $V(t)^2$ is equal to $V_{DC}^2$, i.e., if $V_{DC}$ is the root-mean-square value of $V(t)$.

The operation of the MEMS voltage reference is based on the maximum of the voltage vs. displacement curve of Fig. 3. This maximum, known as the pull-in voltage $V_{pi}$, occurs at $x = d/3$ and has a value

$$V_{pi} = \sqrt{\frac{8kd^3}{27\varepsilon A}} \quad . \tag{3}$$

When the component is biased near the pull-in point, the voltage is insensitive to small changes of $x$. The value of the voltage, $V_{pi}$, is very stable due to the excellent mechanical properties of single-crystal silicon. It depends only on geometry, modulus of elasticity of silicon, and permittivity of the medium.

If the component is voltage biased, it becomes unstable when $V$ exceeds $V_{pi}$ (or $x$ exceeds $d/3$) and the moving electrode will cling to the fixed one. However, if charge rather than voltage is controlled, the system remains stable at the pull-in point $x = d/3$ and beyond. This follows from the fact that the electrical force between the plates of a capacitor with fixed charge $q$ is

$$F_{el,q} = \frac{q^2}{2\varepsilon A} \quad , \tag{4}$$

which does not increase when the distance between the plates ($d - x$) decreases. This should be compared with Eq. (1), which shows that the electrical force in a voltage biased capacitor diverges when the plates approach each other. The relation between charge and voltage of a moving-plate capacitor is obtained by combining Eq. (2) with the force-balance equation between $F_{el,q}$ and spring force $F = -kx$. The result is

$$V = \frac{q}{\varepsilon A}\left(d - \frac{q^2}{2\varepsilon Ak}\right) \quad , \tag{5}$$

which has a maximum $V = V_{pi}$ at $q = q_{pi} \equiv (2\varepsilon Akd/3)^{1/2}$. Charge control requires complicated feedback electronics at DC, but at AC it can be implemented simply by using current bias instead of voltage bias for the drive signal.

## 3   DC and low-frequency applications

Several applications of capacitive MEMS components in electrical metrology have been proposed in Ref. [1]. These include, for example, a voltage divider for high voltages and a current reference. In this article, we concentrate on those applications which have been investigated most extensively: MEMS-based voltage reference and RMS-to-DC converter, which are the topic of this Chapter, and the high-frequency power sensor, which will be discussed in the following Chapter.

Voltage references are fundamental building blocks in a variety of instruments operating over a wide range of accuracies, such as data logging systems and digital multimeters. Calibration laboratories have special voltage reference instruments to maintain a local standard for voltage, and lower level equipment is calibrated against these instruments. The key performance criterions for a voltage reference are the stability with time and temperature, in particular the predictability of temporal drift, immunity to effects of humidity and atmospheric pressure, and immunity to effects of transportation in some applications.

The most accurate and absolute method to generate DC voltages is a Josephson voltage standard (JVS), which can reach relative uncertainty of about $10^{-10}$ at voltage levels between 1 V and 10 V. Such devices are used as the primary voltage standard in most National Standards Laboratories. The most stable DC voltage references with commercial potential are Zener diodes. Temperature stabilised Zener diodes are presently indispensable in high-end multimeters, multifunction calibrators, and DC voltage sources and standards. Long-term stability in the range of about $10^{-6}$/year can be achieved. The main weaknesses of Zener diode references are their dominant 1/$f$ noise component, temperature hysteresis requiring continuous temperature control, dependence on absolute pressure, and considerable power consumption which limits the convenience of portable operation.

MEMS voltage references can provide a competitive alternative to the present voltage references in practically all instruments, even in the most demanding metrology applications, which is the focus of this article [13]. Advantages of MEMS voltage references are low price in mass production, stability, low 1/$f$ noise compared to Zener diodes because of the larger size of the active element, possibility to integrate electronics, tunability over a wide dynamic range (0.1 V - 100 V) by changing the dimensions, and low power consumption. Many of these advantages are still waiting for a practical realisation, though. For example, slow electrostatic charging effects limit the drift rate of the MEMS DC reference to much higher level than expected according to mechanical properties of silicon. However, in Section 3.2 we show that this problem does not exist in a properly designed AC voltage reference.

Another application of electrostatic MEMS components in low-frequency electrical metrology and instrumentation is RMS-to-DC conversion. Measurement of AC voltage is usually done by converting the signal to DC. In the simplest instruments, AC-to-DC conversion is done by simple rectifier circuits, which actually detect the peak value. More advanced methods, such as analog multiplication or fast sampling combined with digital signal processing, are needed to measure the true RMS value of arbitrary waveforms. In the most demanding metrological applications, thermal converters are used: RMS value of AC voltage (or current) is obtained by determining the DC value needed to cause the same heating effect. Most accurate devices are micromachined planar multijunction thermal converters (PMJTC), which can reach relative AC/DC difference below $10^{-6}$ in audio frequency range [18].

Electrostatic MEMS components offer a new solution to true RMS-to-DC conversion. Some of their advantages are small size, low power consumption, short response time, and inherently linear operation. Capacive MEMS-based RMS-to-DC converter has a very high input impedance, which means that its loading effect is much lower than that of the thermal converters. A major drawback is that its operation range cannot easily be extended down to 10 Hz or 100 Hz frequencies without compromising noise properties. The reason for this is the mechanical resonance of the MEMS structure. On the other hand, at high frequencies the bandwidth can extend into the GHz range. Experiments have demonstrated the feasibility of RMS-to-DC conversion with capacitive MEMS structures, but the components are not yet good enough for applications.

## 3.1 DC voltage reference

### 3.1.1 Pull-in voltage and its stability at DC

Use of MEMS as a voltage reference is based on keeping the moving plate of the component at or near the the pull-in point ($x = d/3$ in Fig. 3). The position of the plate can be monitored capacitively. Figure 4 shows the capacitance-voltage curve of a typical voltage-biased component. When the bias voltage with either positive or negative polarity is increased, the gap between the moving and fixed electrodes decreases, and the capacitance increases. This continues until the pull-in voltage $V_{pi}^+$ or $V_{pi}^-$ is reached and the moving electrode hits the mechanical stopper which prevents direct contact between the electrodes.

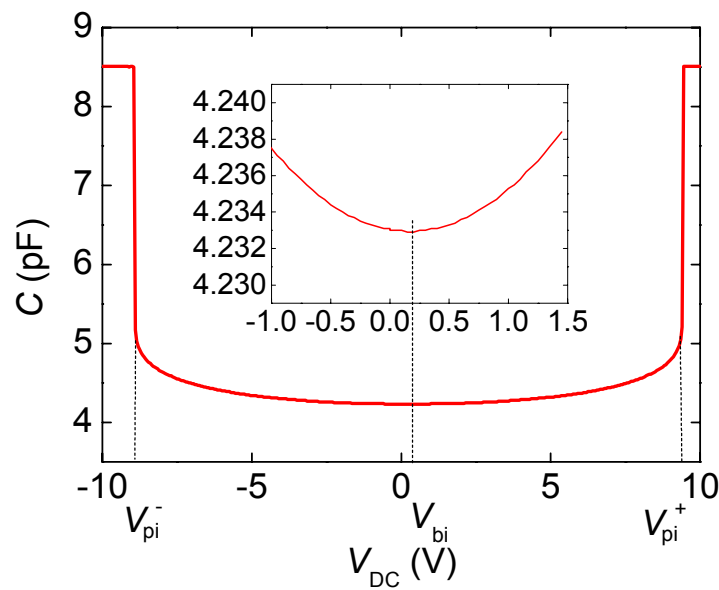

*Figure 4 -* Measured *C-V* curve of a micromechanical moving-plate capacitor.

Note that the symmetry axis of the capacitance-voltage curve is not at $V = 0$ but at $V \approx 0.2$ V, and $V_{pi}^- \neq -V_{pi}^+$. This offset is mainly caused by the built-in voltage $V_{bi}$, which originates from the work function differences between the materials of the component. Charges on the surfaces and interfaces and in the dielectric layers, such as silicon oxide, contribute to the offset voltage, as well. The effective voltage which causes the electrostatic force between the electrodes of the moving capacitor can be written as $V = V_{bias} + V_{bi} + V_c$, where $V_{bias}$ is the externally applied bias voltage and $V_c$ describes the effect of charges in the oxide layers and surfaces.

Slow changes and fluctuations in charge distribution are a major factor limiting the electrical stability of the component. Especially, change of the bias voltage causes slow charging or discharging of the trapping sites, for example, which is observed as a slow drift in the apparent built-in voltage ($V_{pi}^+$ or $V_{pi}^-$). Charging problems can be decreased but not totally eliminated by metallising the electrodes. Slowly changing surface potentials due to, e.g., adsorption and desorption of gas molecules, as well as temperature dependence of the built-in voltage, also contribute to electrical instability.

Stability of the pull-in voltage can be determined, for example, by biasing the MEMS component with a constant DC voltage just below the pull-in value and monitoring the time-dependence of capacitance. Change of pull-in voltage, $\Delta V_{pi}$, is obtained from the measured variation of capacitance, $\Delta C$, as $\Delta V_{pi} = (dC/dV)^{-1}\Delta C$, where $dC/dV$ is the derivative of the *C-V* curve at the measurement point. MIKES and VTT used this method in the EMMA project to determine the long-term stability of $V_{pi}$ in several different types of moving-plate microstructures. When the bias voltage was increased from 0 to the final value near the pull-in point, the drift of $\Delta V_{pi}$ was rapid during the initial period which lasted between about 1 and 10 hours in different samples. After this, the drift rate decreased and remained more or less constant during the measurement, which was continued for several days. The final drift rate in terms of relative variation of the pull-in voltage was of the order of 50 µV/V in day (24 hours) even in the most stable samples. Most reproducible results were obtained with the bulk micromachined structure shown in Fig. 2b.

Long-term stability of the pull-in voltage has also been studied by the Delft group [19], [20]. Their MEMS structures are fabricated using the epi-poly process, which is one variation of surface micromachining. Their results show that during the first week after starting the experiment, the built-in voltage decreases by about 0.1% or 0.2%. Then $V_{pi}$ stabilises and does not vary more than about 100 µV/V [19] or 500 µV/V [20] during the following two or three weeks.

The slow drift observed in effective pull-in voltage after applying a DC voltage bias is mainly caused by electrostatic charging effects. Another important factor which affects the stability of pull-in voltage is its temperature dependence. The temperature coefficient of elastic modulus of single-crystal silicon causes a temperature dependence of about $-6 \cdot 10^{-5}$/°C in the pull-in voltage, but other factors such as component mounting can sometimes cause a larger temperature dependence. In any case, temperature dependence is so strong that the MEMS component must be temperature stabilised in metrological applications. An open capacitive structure is also very sensitive to changes in humidity, pressure, etc., and the component must be encapsulated in vacuum or in a protective gas atmosphere.

### 3.1.2 Feedback-controlled voltage reference

Figure 5 presents the block diagram of control electronics of a complete MEMS-based DC voltage reference [17]. The device is based on a MEMS with elastically supported seesaw-type electrode (see Fig. 2c), which is symmetrically positioned above six rigid electrodes to form six capacitances. The seesaw is stabilised to the pull-in position by controlling the DC voltage on the two rightmost electrodes (1, 2). These electrodes have equal areas and are in practice equidistant from the silicon springs. The idea is to apply DC voltage with opposite polarities but with slightly different magnitudes to the two electrodes. The direction of the electrostatic force does not depend on the polarity of the voltage, but in this way the built-in voltage can be compensated and the effect of its small changes can be eliminated. The position of the seesaw electrode is monitored by the AC capacitive bridge formed of innermost electrodes (3, 4). The ratio of the AC voltage levels applied to those electrodes with opposite polarity is adjusted so that the voltage $V$ of the seesaw vanishes when the seesaw is at pull-in position. Deflection from the pull-in position causes an AC voltage $V$, which is used as a feedback signal and mixed with a reference voltage $V_{ext}$. The DC component of the mixer output is fed to a $PI^{3/2}$ controller in the feedback loop which restores the seesaw electrode back to the pull-in position. The controller output can be used as the DC reference voltage.

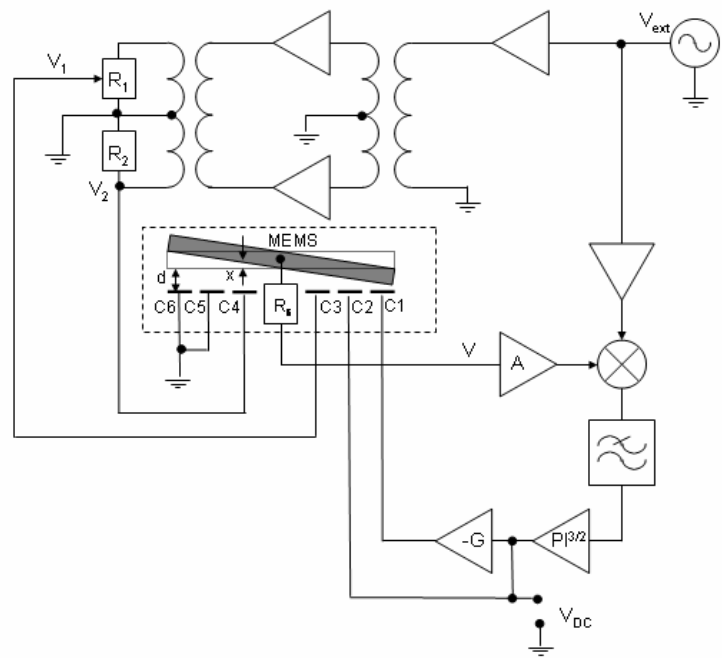

*Figure 5 -* Block diagram of a DC voltage reference based on feedback control. Modified from Ref. [17].

The DC voltage reference based on the seesaw type SOI component of Fig. 2c and the control electronics of Fig. 5 has been fabricated and tested [17]. In an 11-hour measurement, the output voltage of the device stayed constant within about 60 µV/V after the initial stabilisation period of about 1 hour. The result is not yet good enough for metrological applications. Further improvements are still needed at least for component packaging and feedback electronics.

## 3.2 AC voltage reference

Many problems of MEMS-based DC voltage references can be solved by using AC actuation instead of DC. Most problems of slow charging effects are eliminated when the high constant electric field (about 10 MV/m) caused by the constant DC bias does not exist between the electrodes [21]. Also, the AC voltage reference can be kept at the pull-in position simply by biasing it with an AC current, and complicated feedback control is not needed. At present, AC voltage references are not used in practical instruments and do not actually exist, but they could replace DC voltage references in many applications.

As was discussed earlier, the results of Section 2.2 are valid not only for DC voltage but also for the root-mean-square value of the AC voltage. This assumes static displacement of the moving electrode. The assumption is best valid at frequencies much higher than the mechanical resonance frequency of the component, which is typically of the order of 5 kHz. Discussion of charge control is valid, as well, if we replace $q$ in Eqs. (4) and (5) by effective "root-mean-square charge" $q_{eq} = I_{RMS}/\omega$, where $I_{RMS}$ is the root-mean-square value of the AC current drive and $\omega$ is the angular frequency of AC current. Thus, with a simple AC current bias we can control the position of the moving plate on both sides of the pull-in maximum of Fig. 3. Pull-in voltage $V_{pi}$ can be determined simply by finding the maximum of voltage as a function of the magnitude of AC current drive. According to Eq. (5) and discussion below it, pull-in point is reached at $I_{RMS} = q_{pi}\omega \equiv (2\varepsilon Akd/3)^{1/2}\omega$.

The component experiences an effective DC voltage even when a pure AC drive is used. Built-in voltage still exists and the charges in the oxide layers and surfaces can give their own, possibly time-dependent contribution. The electrical force affecting the moving plate is determined by total RMS voltage, $(V_{AC}^2 + V_{DC}^2)^{1/2}$. Here $V_{AC}$ is the AC voltage caused by the current drive and $V_{DC}$ is the sum of all internal and external DC voltages affecting the MEMS component. In this case, the maximum RMS value of the AC voltage as a function of AC current drive is

$$V_{AC,max} = \sqrt{V_{pi}^2 - V_{DC}^2} \qquad (6)$$

instead of $V_{pi}$. If $V_{DC}$ remained stable with time, it would not jeopardise the stability of the device. However, in practice $V_{DC}$ changes with time due to charging effects, for example, and those changes can be observed at $V_{AC,max}$.

The effect of internal DC voltages on the stability of the MEMS-based AC voltage reference can be eliminated by compensating the built-in voltage with an applied DC voltage [12]. Small variations of $V_{DC}$ with time, $\Delta V_{DC}$, can be described by writing $V_{DC} = V_{DC,0} + \Delta V_{DC}$, where $V_{DC,0}$ is a constant. Because $\Delta V_{DC} \ll V_{DC,0} \ll V_{pi}$, we can write Eq. (6) in the form

$$V_{AC,max} = V_{pi} - \frac{V_{DC,0}^2}{2V_{pi}} - \frac{V_{DC,0}}{V_{pi}}\Delta V_{DC} \,. \qquad (7)$$

This shows that when the built-in voltage and other internal DC voltages of the component are compensated by an applied constant DC voltage in such a way that $V_{DC,0} = 0$, small changes of the DC voltage do not affect the reference voltage $V_{AC,max}$ in the first order.

The properties of AC voltage reference without built-in voltage compensation are reported in Ref. [22]. Here we concentrate on the results obtained with compensation of the internal DC voltages [12]. Simple control electronics was constructed to provide AC current bias and DC voltage compensation for the bulk-micromachined component, whose structure is shown in Fig. 2b. The stability measurement was made by recording the RMS value of the AC component of the MEMS voltage, $V_{AC}$, while the amplitude of the driving 100 kHz current was repeatedly varied across the maximum of the output voltage, $V_{AC,max}$ (related to pull-in voltage, about 9 V for this sample, by Eq. (6)). The value of $V_{AC,max}$ was determined from a parabolic fit made to the data. When the built-in voltage compensation was not used, the drift rate of $V_{AC,max}$ was about -2 µV/V per day. The drift rate could be controlled by the applied DC bias voltage, $V_{bias}$, and even the sign of the drift could be changed. As Eq. (7) predicts, the dependence between the drift rate and $V_{bias}$ was roughly linear.

As expected, the drift rate was minimised when $V_{bias}$ was selected to be close to the measured C-V curve minimum ($V_{bi}$ in Fig. 4), with opposite polarity. Stability measurement with this bias voltage was continued for more than three weeks. The results are plotted in Fig. 6 after correcting the data for changes in ambient pressure (-6 µV/hPa) and temperature (-180 µV/°C). No drift can be observed in the 100 kHz output voltage of the device within the measurement accuracy of about 2 µV/V during the 3-week period. After the stability test was finished, both the AC drive and the DC bias (built-in voltage compensation) were disconnected. After one day the voltages were reconnected and the measurement continued. After a relaxation period of about 8 hours, the output voltage returned to the original level within the measurement accuracy of about 2 µV/V [12].

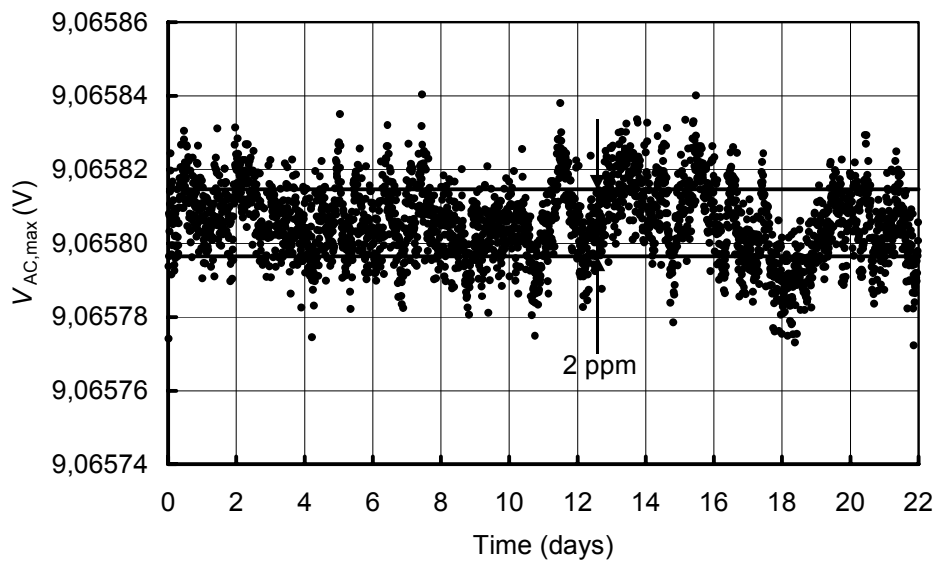

*Figure 6 -* Time stability of the 100 kHz AC voltage reference with built-in voltage compensation. Modified from Ref. [12].

The measured drift rate below 2 µV/V in three weeks is the state-of-the-art of long term stability of MEMS devices based on the pull-in voltage, and is already sufficient even for metrological applications.

### 3.3 True RMS-to-DC converter

Measurement of the balance between forces arising from AC and DC voltages is the most obvious application of moving-plate MEMS in electrical metrology. As we discussed in Section 2.2, operation of MEMS as a true RMS-to-DC converter is based on the quadratic voltage dependence of electrostatic force and on static displacement of the moving plate. The best way to ensure static displacement is to make sure that the frequency of the AC voltage is much higher than the mechanical resonance frequency of the component. This limits the lowest operating frequency of the device quite seriously, because resonance frequencies of the MEMS components are typically in the kHz range. Resonance frequency can be lowered by decreasing the spring constant or by increasing the mass of the moving plate, but this makes the system more susceptible to mechanical vibrations. It is also possible to tune the spring constant electrically [1]. RMS-to-DC converter can be used below the resonance frequency of the component by damping the mechanical oscillations by means of increased dissipation, but this increases the effect of mechanical noise. A better way would be to use active damping with electrical feedback [1].

Micromachined electrostatic true RMS-to-DC converter was first proposed by van Drieënhuizen and Wolffenbuttel in 1995 [3]. Their approach was based on two surface micromachined capacitors, both of which had two fixed electrodes and a moving electrode in the middle. One of the components was actuated by connecting the input AC voltage between the bottom electrode and the moving middle electrode, and the other component was actuated similarly with a DC voltage. Top and middle electrodes were used to measure the difference of deflection of the moving plate with a capacitive bridge method. The RMS value of the AC voltage was determined by measuring the DC voltage which caused the same deflection as the AC voltage. Squeeze film damping of air between the plates was used to eliminate the vibrations of the moving electrode. Measurement results demonstrated the feasibility of the method [5], but voltage-to-displacement sensitivity turned out to be too small for practical implementation. The sensitivity could be improved by a bulk micromachined device, in which the vibrations were damped by means of perforation of the moving electrode, but only DC tests of that device have been published [6].

Another attempt to realize a MEMS-based RMS-to-DC converter has been made by the VTT group [7]. Their structure is based on a micromechanical seesaw electrode between two fixed plates. Both of the fixed plates have separate electrodes on both sides of the pivot point of the seesaw. The measured AC voltage is applied between the seesaw and, say, the left electrode on the top plate. A DC voltage is then applied between the seesaw and the left electrode on the bottom plate, so that the electrical forces tend to turn the seesaw to different directions. The position of the seesaw is measured capacitively using the electrodes on the right, and the DC voltage is adjusted so that the forces of AC and DC voltages cancel each other. SOI-based components for this purpose have been designed, but the actual test was made with a commercial micromachined seesaw-type gyroscope. The operation principle was demonstrated, but spurious charges on silicon dioxide surfaces and asymmetrical capacitance-voltage characteristics prevented accurate measurements [7]. After these preliminary measurements, the work on metrology applications of MEMS at VTT has concentrated on developing voltage references and high frequency power sensors, and the problems with RMS-to-DC converter are waiting for solutions.

# 4 High frequency power sensors

Power detection is an important part of high frequency (hf) applications and measurement technology. Existing power meters are designed to be used in test and measurement applications as the last instrument in the measurement path, using a power absorbing terminating sensor. Three types of commercial sensors exist for radio frequency and microwave power measurements: thermistors, thermocouples, and diode detectors. Thermistors and thermocouples are based on converting the hf power to heat and measuring the temperature. In diode detectors, power measurement is based on nonlinear current-voltage characteristics of fast Schottky diodes: at low power levels, the diodes respond to the RMS value of the signal and produce a DC output proportional to the applied hf power. Thermistors are widely used as power transfer standards in calibration laboratories, but in many other applications they have been replaced by the more sensitive thermocouples. Diode detectors are fast and offer measurement capability at power levels below the detection limit of the thermal sensors, down to about -70 dBm (100 pW), but diode detectors are nonlinear at power levels above about -20 dBm (10 µW).

MEMS technology offers a new alternative to high frequency power measurements. The measurement principle is the same as in lower frequency RMS-to-DC converters (see Section 3.3). In principle, the same sensor could be used for power measurements in a wide frequency range extending from 10 kHz and below to 40 GHz and above. As opposed to conventional dissipating and terminating hf power detectors, the MEMS device is intrinsically a "through sensor", i.e., it is connected in the middle of the signal path to measure the power that is transmitted through the sensor. This property opens up new possibilities in, e.g., level sensing and control applications for signal sources and transmission. Terminating measurement, which has advantages in some signal measurement applications, can be realised with the MEMS sensor combined with a termination.

## 4.1 Design considerations

A schematic view of a seesaw-type high frequency power sensor is presented in Fig. 7. It has a similar structure as the low frequency RMS-to-DC converter, except that the bottom electrode has been replaced by a transmission line, in this case by a coplanar waveguide (CPW). The hf signal in the transmission line induces an electrostatic attractive force which tends to tilt the seesaw plate. According to Eq. (1), the force depends quadratically on the hf voltage, i.e., it is directly proportional to hf signal power in the line. The movement of the seesaw plate is detected capacitively using the sense electrodes on both sides of the torsional spring of the seesaw. In order to increase the dynamic range of the sensor, the tilt can be compensated by applying a DC voltage to the force feedback electrode at the opposite end of the seesaw.

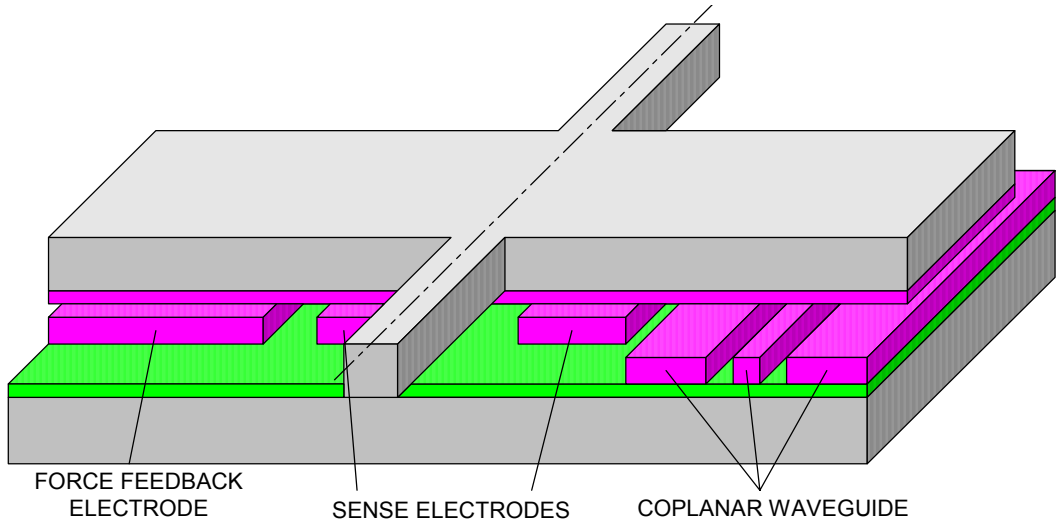

*Figure 7 -* Schematic view of a seesaw-type high frequency power sensor (modified from Ref. [23]).

A simplified model of the MEMS power sensor is a capacitor $C$ embedded in the middle of a transmission line with characteristic impedance $Z_0$ (typically 50 Ω), as shown in Fig. 8a. The sensor responds to the average of the square of the voltage, $V_{RMS}^2$, and the hf power is obtained as $P_{hf} = V_{RMS}^2/Z_0$ using the known value of $Z_0$. Power-sensing capacitance $C$ between the signal line and the moving electrode (seesaw in Fig. 7) causes a discontinuity in the transmission line and inevitably gives rise to reflections and transmission losses. Their minimisation is an important design issue. Note, however, that in an ideal transmission line without any discontinuities there would not exist any net forces: the attractive electrostatic force would be exactly cancelled by the repulsive magnetic force.

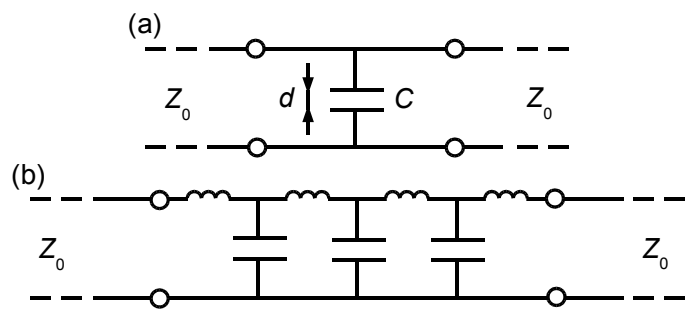

*Figure 8 -* (a) Simplified model of a MEMS power sensor in a transmission line. (b) Increasing the frequency band by adding inductive elements and dividing the power-sensing capacitor into many parallel components.

The power resolution of the sensor is fundamentally limited by the mechanical Brownian noise force density of the plate resonator, $f_{m,n} = (4k_BT\gamma)^{1/2}$ (in N/Hz$^{1/2}$), where $k_B$ is the Boltzmann constant, $T$ is the absolute temperature, $\gamma = k/(\omega_r Q)$ is the mechanical damping factor, and $k$, $\omega_r$ and $Q$ are the spring constant, quality factor and resonance frequency of the mechanical resonator, respectively [11]. Ultimate power density resolution, $p_{res}$ (in W/Hz$^{1/2}$), is obtained by equating the noise force with the electrostatic force given by Eq. (1), and the result is

$$p_{\text{res}} = \frac{2d^2}{Z_0 \varepsilon A} \sqrt{4 k_B T \gamma} \ . \tag{8}$$

Here we have assumed that $x = 0$ in Eq. (1), i.e., that the displacement of the plate is compensated by a DC voltage. Equation (8) shows that sensitivity can be improved by decreasing the gap $d$ of the sensing capacitor, but in practice the gap cannot be made much smaller than about 0.5 µm. Moreover, the RF bandwidth of the sensor, $\Delta\omega = 2/(Z_0 C) = 2d/(Z_0 \varepsilon A)$, decreases when $d$ is decreased, so there is a tradeoff between resolution and bandwidth. Maximum hf power determined by the pull-in of the component decreases with decreasing gap thickness, as well [11].

By proper matching, the bandwidth of the sensor can be effectively increased without compromising the sensitivity. A simple way to improve high frequency properties is to add inductive elements into the signal line on both sides of the moving-plate capacitor [23]. Even better results are obtained by splitting the sensing capacitor to many parallel components and adding inductive elements in between, as shown schematically in Fig. 8b. In practice this can be done by splitting the seesaw electrode into a fork-like structure above the transmission line [11]. In this way, a SOI-based MEMS hf power sensor has been designed with simulated power resolution of -50 dBm (10 nW) and the reflection coefficient, or the scattering parameter $S_{11}$, smaller than -27 dB at frequencies below 40 GHz. The simulated value of the $S_{21}$ scattering parameter, which describes the attenuation of the signal when it is transmitted through the sensor, is better than -0.1 dB for the design with 4 GHz bandwidth and better than -1 dB for the design with 40 GHz bandwidth [11]. Note, however, that the calculated power resolution is based on the assumption that the noise of the measuring electronics can be decreased below the mechanical Brownian noise level. In practice this is a challenging task.

## 4.2 Experimentally tested devices

The seesaw-type hf power sensor described in the previous section was designed to be a SOI-based structure. VTT has developed a new process for fabricating such components and has had continuous progress, but so far there have been no properly operating sensors available for measurements. Somewhat simpler SOI-based hf power sensors, so-called railroad sensors, have been fabricated and tested by VTT, but with that structure the transmission losses were high, bandwidth was limited to 5 GHz, and achievable power resolution was too poor [24]. Most of the experimental tests of the MEMS hf power sensors have been performed using surface micromachined components. Their ultimate stability and sensitivity properties are not expected to be as good as those of the optimized SOI components, but development of processes to fabricate operating components has been faster.

Researchers in the University of Twente have developed several versions of a surface micromachined high frequency power sensors. A microscope picture of their latest sensor is presented in Fig. 9a [10]. The sensor consists of a coplanar waveguide, above which there is an aluminium membrane in a bridge-like manner. The motion of the membrane is detected capacitively. Substrate material is low-loss AF45 glass. Measured change of capacitance as a function of power of a 1 GHz signal is presented in Fig. 9b for two sizes of the sensing membrane. For the larger membrane, a sensitivity of 180 fF/W is obtained. This corresponds to a resolution of about -3 dBm (0.5 mW), limited by the noise in the capacitance measurements. This is still far from the theoretical noise limit, which is estimated to be in the range of several nanowatts. The scattering parameters were measured up to 4 GHz, with results $S_{11}$ < -30 dB and $S_{21}$ > -0.15 dB.

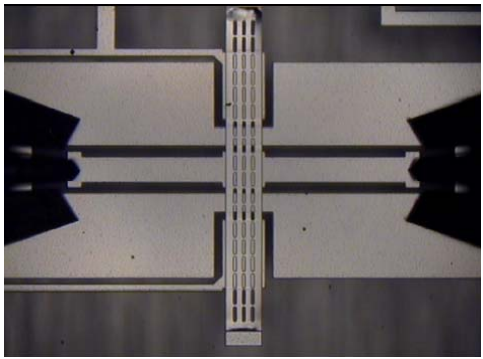 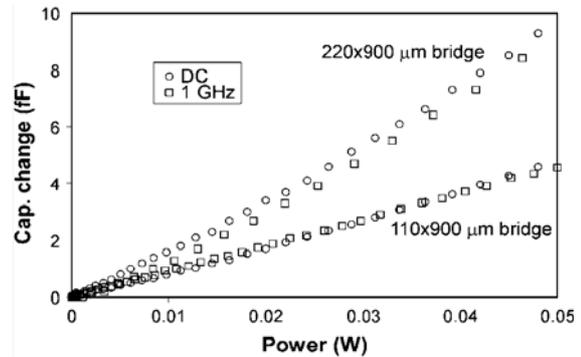

Figure 9 -   (a) Microscope picture of a surface micromachined power sensor fabricated in the University of Twente [10].  (b) Measured capacitance change on 220 × 900 µm² and 110 × 900 µm² bridge sensors as a function of power for a 1 GHz signal and DC actuation [10].

VTT has fabricated and tested surface micromachined hf power sensors, as well [25]. The structure of the sensor was quite similar to the bridge power sensor of Fig. 9a. The component was fabricated on a quartz substrate using Mo-Si-N amorphous metal micromachining. Scattering parameters were measured up to 50 GHz, indicating that the reflection coefficient $S_{11}$ was lower than -28 dB at frequencies below 40 GHz, and $S_{21}$ was better than -0.6 dB in the whole frequency range up to 50 GHz. Capacitance measurements of the moving membrane indicated roughly linear power dependence, with sensitivity between 16 fF/W and 180 fF/W for 26 GHz signal, depending on the DC bias: the best sensitivity was achieved when the DC bias was near the pull-in voltage.

## 5   MEMS reference oscillators

There are several microsystem technologies that can be used to realize resonators in the frequency range ranging from a few kHz up to several GHz. The relevant technologies include surface acoustic wave (SAW) and bulk acoustic wave (BAW) technologies, and micromachining of silicon resonators. All these technologies could be used for implementing miniaturized reference oscillators. The driving force for development is the telecommunication industry. BAW technology has already been industrialized for RF filters of mobile phones. According to market surveys and technology foresights, the next major microsystem component in the RF market will be a MEMS resonator for the reference oscillator as a substitute to the quartz crystal resonator used presently. Several companies, such as Discera, SiTime and VTI Technologies, are trying to open the market for MEMS-based oscillators.

Long-term stability is the key property of a good reference oscillator. The benchmarking results on this issue have been published by Kaajakari et al [26]. These authors fabricated micromachined single-crystalline, square-shaped silicon resonators on a SOI wafer. The resonance mode of the devices was the so-called square-extensional mode at about 13 MHz. The resonators were packaged at a wafer level in a low-pressure atmosphere less than 1 mbar. Long-term stability of the resonators, placed in a temperature-controlled chamber, was investigated by monitoring the RF transmission curves. According to the data, reproduced in Fig. 10, the relative stability of the resonance frequency is much better than $10^{-7}$/month with no observable drift. This level of stability is comparable to that of high-quality quartz crystals.

Kaajakari and coworkers investigated also the role of packaging-related effects, such as stress relaxation, by performing measurements using different resonator geometries. The conclusion drawn by the authors was that package-induced stresses can propagate into the MEMS component through mechanical anchors. Therefore the design of the anchoring structures is crucial for achieving highly stable MEMS components.

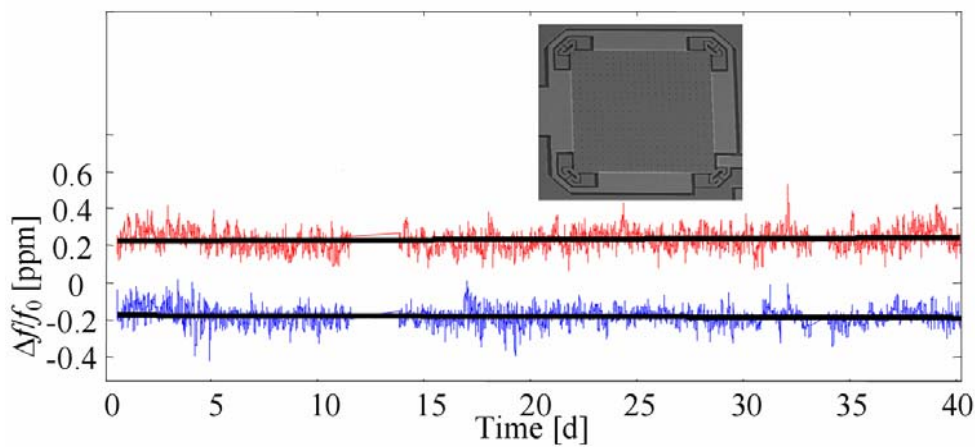

*Figure 10 -* Stability of the mechanical resonance frequencies of two 13-MHz square-extentional mode resonators (blue and red traces, respectively) over 40 days. Vertical axis shows relative frequency change in parts per million (ppm), and horizontal axis shows time in days. Inset: photograph of the resonator.

# 6  Conclusions

MEMS-based devices have a great potential in electrical metrology and instrumentation. They compare favourably to the existing devices by size, price, and power consumption. Operation principle of balance between electrical and mechanical forces in capacitive moving-plate MEMS components can be applied, for example, in DC and AC voltage references, RMS-to-DC converters, and high frequency power sensors. The ultimate stability of the components fabricated from single-crystal silicon is expected to be very good because of the excellent mechanical properties of single crystal silicon. A low 1/*f* noise is expected, as well. In practice, however, the stability of the MEMS components is often limited by electrostatic charging effects on the surfaces and interfaces and in dielectric layers. Hermetic encapsulation and temperature control of the components is required due to the strong temperature and pressure dependencies, but mechanical stresses caused by packaging are a potential source of problems. Being mechanical devices, the moving-plate MEMS components are inherently sensitive to mechanical vibrations, but their effect can be minimised with a proper design.

MEMS devices for metrological applications have been fabricated using different processing technologies: bulk micromachining, surface micromachining, and the silicon-on-insulator (SOI) process. Prototypes of RMS-to-DC converters have been fabricated and tested, but their properties are not yet good enough for real applications. In recent years, development of MEMS devices for electrical low-frequency applications has mainly focused on DC and AC voltage references based on the pull-in effect of micromechanical moving-plate capacitors. Due to electrostatic charging effects, DC voltage references are not yet stable enough for metrology. The problems can be largely circumvented by using AC actuation instead of DC and by compensating the internal DC voltages of the MEMS component. In this way, an AC voltage reference with relative voltage drift smaller than $2 \cdot 10^{-6}$ during a three-week measurement period has been demonstrated. This level of stability is already good enough for many applications even in metrology. Development of a DC reference with a similar or better stability still remains a challenge, even though the DC reference could be replaced by an AC reference in many instruments.

Development of high frequency power sensors was another focus area of the EMMA collaboration project, which has produced most of the results summarised in this article. The MEMS device measures intrinsically the RF or microwave power transmitted through the sensor. This principle opens up new possibilities compared to conventional methods, in which the power absorbed in the sensor is measured. Several types of components have been designed for frequencies up to 4 GHz or 40 GHz. Operation principle has been demonstrated in several test devices, but their noise level is still far above the theoretical limit. Prospects of MEMS-based high frequency power sensors are very promising, but a lot of further research and development is needed before their signal-to-noise ratio and other properties are at level required in practical applications.

MEMS components have also been developed for use as stable frequency references. Their main advantage compared to the conventional quartz crystals is small size, which makes the MEMS resonators especially attractive for mobile applications. The measured drift of resonance frequency of a SOI-based square-extensional plate resonator is not more than about $10^{-8}$/month. This result is important in that it represents the best so-far-demonstrated stability for any silicon-based MEMS components. If this level of stability could be achieved also for the MEMS voltage references, Zener references would be replaced by MEMS voltage references in high-end applications. In the MEMS-based voltage references, however, the situation is more demanding in that the stability of the reference depends both on mechanical and electrical properties of the device, whereas the stability of the MEMS resonators depends primarily on mechanical properties.

# References

[1] H. Seppä, J. Kyynäräinen, and A. Oja, Microelectromechanical Systems in Electrical Metrology. IEEE Trans. Instrum. Meas. 50 (2001), pp. 440 - 444.
[2] R.F. Wolffenbuttel, and C.J. van Mullem, The Relationship Between Microsystem Technology and Metrology. IEEE Trans. Instrum. Meas. 50 (2001),

pp. 1469 - 1474.
[3] B.P. van Drieënhuizen and R.F. Wolffenbuttel, Integrated Micromachined Electrostatic True RMS-to-DC Converter. IEEE Trans. Instrum. Meas. 44 (1995), pp. 370 - 373.
[4] M. Suhonen, H. Seppä, A.S. Oja, M. Heinilä, and I. Näkki, AC and DC Voltage Standards Based on Silicon Micromechanics. CPEM 98 Digests, Washington DC, 1998, pp. 23 - 24.
[5] B.P. van Drieënhuizen, Integrated Electrostatic RMS-to-DC Converter. Ph.D. Thesis, Delft University, 1996.
[6] G. de Graaf, M. Bartek, Z. Xiao, C.J. van Mullem, and R.F. Wolffenbuttel, Bulk Micromachined Electrostatic True RMS-to-DC Converter. IEEE Trans. Instrum. Meas. 50 (2001), pp. 1508 - 1512.
[7] J. Kyynäräinen, A.S. Oja, and H. Seppä, A Micromechanical RMS-to-DC Converter. CPEM 2000 Digests, Sydney, 2000, pp. 699 - 700.
[8] J. Kyynäräinen, A.S. Oja, and H. Seppä, Stability of Microelectro-mechanical Devices for Electrical Metrology. IEEE Trans. Instrum. Meas. 50 (2001), pp. 1499 - 1503.
[9] E. Cretu, L.A. Rocha, and R.F. Wolffenbuttel, Micromechanical Voltage Reference Using the Pull-in of a Beam. IEEE Trans. Instrum. Meas. 50 (2001), pp. 1504 - 1507.
[10] L.J. Fernández, R.J. Wiegerink, J. Flokstra, J. Sesé, H.V. Jansen, and M. Elwenspoek, A Capacitive RF Power Sensor Based on MEMS Technology. J. Micromech. Microeng. 16 (2006), pp. 1099 - 1107.
[11] A. Alastalo, J. Kyynäräinen, H. Seppä, A. Kärkkäinen, N. Pesonen, M. Lahdes, T. Vähä-Heikkilä, P. Pekko, and J. Dekker, Wideband Microwave Power Sensor Based on MEMS Technology. CPEM 2004 Digests, London, 2004, pp. 115 - 116.
[12] A. Kärkkäinen, N. Tisnek, A. Manninen, N. Pesonen, A. Oja, and H. Seppä, Electrical Stability of a MEMS-Based AC Voltage Reference. Sensors and Actuators, submitted (2006).
[13] A. Kärkkäinen, MEMS Based Voltage References. Ph.D. Thesis, Helsinki University of Technology, 2006.
[14] S. Franssila, Introduction to Microfabrication, John Wiley & Sons, 2004.
[15] A. Kärkkäinen, P. Pekko, J. Dekker, N. Pesonen, M. Suhonen, A. Oja, J. Kyynäräinen, and H. Seppä, Stable SOI Micromachined Electrostatic AC Voltage Reference. Microsyst. Technol. 12 (2005), pp. 169 - 172.
[16] www.vti.fi.
[17] A. Kärkkäinen, S.A. Awan, J. Kyynäräinen, P. Pekko, A.S. Oja, and H. Seppä, Optimized Design and Process for Making a DC Voltage Reference Based on MEMS. IEEE Trans. Instrum. Meas. 54 (2005), pp. 563 - 566.
[18] M. Klonz, H. Laiz, and E. Kessler, Development of Thin-Film Multijunction Thermal Converters at PTB/IPHT, IEEE Trans. Instr. Meas. 50 (2001), pp. 1490 - 1498.
[19] L.A. Rocha, E. Cretu, and R.F. Wolffenbuttel, Stability of a Micromechanical Pull-in Voltage Reference. IEEE Trans. Instrum. Meas. 52 (2003), pp. 457 - 460.
[20] L.A. Rocha, E. Cretu, and R.F. Wolffenbuttel, Analysis and Analytical Modeling of Static pull-in with Application to MEMS-Based Voltage reference and Process Monitoring. J. Microelectromech. Syst. 13 (2004), pp. 342 - 354.
[21] A. Kärkkäinen, A.S. Oja, J. Kyynäräinen, H. Kuisma, and H. Seppä, Stability of Electrostatic Actuation of MEMS. Physica Scripta T114 (2004), pp. 193 - 194.
[22] A. Kärkkäinen, N. Pesonen, M. Suhonen, A.S. Oja, A. Manninen, N. Tisnek, and

H. Seppä, MEMS-Based AC Voltage Reference, IEEE Trans. Instrum. Meas. 54 (2005), pp. 595 - 599.

[23] T. Vähä-Heikkilä, J. Kyynäräinen, A. Oja, J. Varis, and H. Seppä, Capacitive MEMS power sensor. Proc. of the 3rd Workshop on MEMS for Millimeterwave Communication, Crete, 2002.

[24] A. Oja, J. Kyynäräinen, A. Kärkkäinen, P. Pekko, J. Dekker, J. Varis, T. Vähä-Heikkilä, and H. Seppä, High-frequency power sensor. Proc. of the 4th Workshop on MEMS for Millimeterwave Communication, Toulouse, 2003.

[25] T. Vähä-Heikkilä, A. Alastalo, J. Kyynäräinen, A. Kärkkäinen, N. Pesonen, M. Lahdes, J. Varis, P. Pekko, J. Dekker, M. Ylönen, H. Kattelus, H. Seppä, and A. Oja, unpublished (2004).

[26] V. Kaajakari, J. Kiihamäki, A. Oja, H. Seppä, S. Pietikäinen, V. Kokkala, and H. Kuisma, Stability of Wafer Level Vacuum Encapsulated Single-Crystal Silicon Resonators. Digest of Technical Papers, Transducers'05, Seoul, 2005, pp. 916 - 919.